\documentclass[twocolumn]{aa}
\usepackage{graphicx}
\usepackage{txfonts}
\begin{document}
   \title{Abundance difference between components of wide binaries
          \thanks{Based on observations made with the Italian Telescopio 
                  Nazionale Galileo (TNG) operated on the island of La Palma
                  by the Centro Galileo Galilei of the INAF (Istituto 
                  Nazionale di Astrofisica) at the Spanish Observatorio del 
                  Roque de los Muchachos of the Instituto de Astrofisica 
                  de Canarias}}

   \author{S. Desidera
          \inst{1},
           R.G. Gratton
          \inst{1},
           S. Scuderi
          \inst{2},
           R.U. Claudi
          \inst{1},
           R. Cosentino,
          \inst{2,3},
           M. Barbieri
          \inst{4},
           G. Bonanno
          \inst{2},
           E. Carretta
          \inst{1},
           M. Endl
          \inst{5},
           S. Lucatello
          \inst{1,6},
           A.F. Martinez Fiorenzano
          \inst{1,6},
          \and
           F. Marzari
          \inst{7}}

   \authorrunning{S. Desidera et al.}

   \offprints{S. Desidera,  \\
              \email{desidera@pd.astro.it} }

   \institute{INAF -- Osservatorio Astronomico di Padova,  
              Vicolo dell' Osservatorio 5, I-35122, Padova, Italy 
             \and
             INAF -- Osservatorio Astrofisico di Catania, Via S.Sofia 78, Catania, Italy
             \and
             INAF -- Centro Galileo Galiei, Calle Alvarez de Abreu 70, 38700 
             Santa Cruz de La Palma (TF), Spain 
             \and
             CISAS -- Universit\'a di Padova, Via Venezia 15, Padova, Italy 
             \and
             McDonald Observatory, The University of Texas at Austin, Austin, 
             TX 78712, USA
             \and 
             Dipartimento di Astronomia -- Universit\'a di Padova, Vicolo
             dell'Osservatorio 2, Padova, Italy 
             \and 
             Dipartimento di Fisica -- Universit\'a di Padova, Via Marzolo 8,
             Padova, Italy }

 \date{Received 16 Febraury 2004 / Accepted 26 Febraury 2004}

   \abstract{We present iron abundance analysis for 23 wide
             binaries with main sequence components in the temperture range
             4900-6300~K, taken from
             the sample of the pairs currently included in the radial
             velocity planet search on going at the Telescopio Nazionale
             Galileo (TNG) using the high resolution spectrograph SARG.
             The use of a line-by-line differential analysis technique 
             between the components of each pair allows us to reach errors
             of about 0.02 dex in the iron content difference.
             Most of the pairs have abundance differences lower than 0.02 dex
             and there are no pairs with differences larger than 
             0.07 dex. The four cases of differences larger than 0.02 dex
             may be spurious because of the larger error bars 
             affecting pairs with large temperature difference, cold stars 
             and rotating stars.
             The pair HD~219542, previously reported by us to have
             a different composition, here is shown to be normal. 
             For non-rotating stars warmer than 5500 K, characterized by a 
             thinner convective envelope and for which our analyis appears
             to be of higher accuracy, we are  able to exclude 
             in most cases the consuption of more than 1 Earth Mass of iron 
             (about 5 Earth masses of meteoritic material) during the main 
             sequence lifetime of the stars, placing more stringent limits
             (about 0.4 Earth masses of iron) in five cases of warm stars.
             This latter limit is similar to the estimates of rocky material
             accreted by the Sun during its main sequence lifetime.
             Combining the results of the present analysis with those for the
             Hyades and Pleiades, we conclude that the hypothesis that
             pollution by planetary material is the only mechanism responsible
             for the highest metallicity of the stars with planets
             may be rejected at more than 99\% level of confidence 
             if the incidence of planets in these samples is as high as 8\%
             and similar to the field stars included in current 
             radial velocity surveys. However,  
             the significance of this result drops
             considerably if the incidence of planets around stars in
             binary systems and
             clusters is less than a half of that around normal field stars.

   \keywords{(Stars:) abundances -- (Stars:) planetary systems -- 
             (Stars:) binaries: visual -- Techniques: spectroscopic}
   }

   \maketitle
%

\section{Introduction}
\label{s:intro}

The evidence for a high metal content in stars harbouring 
planets is becoming stronger as planet discoveries cumulate
and suitable control samples are studied using strictly the same procedures.
It appears that the metal abundance of stars with planets is on average 
about 0.25 dex larger than that of typical control sample stars 
(Laws et al.~\cite{laws03}) and that
the frequency of planets is a strong function of metallicity
(Santos et al.~\cite{santos04}, Fischer et al.~\cite{fischer04}).

Two alternative hypotheses have been proposed to explain these observations:
either the high metallicity is responsible for the presence of planets,
making their formation easier; or the planets are the cause of
the high metallicity, because of pollution of metal-rich planetary material
onto the (outer region of the) central star (Gonzalez \cite{gonzalez}).

Infall of planetesimals on the star during the early phases of planet
formation is generally expected on the basis of current models
of planet formation.
The orbital migration proposed to explain the
occurrence of the close-in giant planets found by
radial velocity surveys also points to the infall on the star
of portions of the proto-planetary disk.

Most of the accretion is expected to take place
during the early phases of the evolution of the planetary system.
However, when a star is still in the phase of
gravitational contraction, its convective zone is much thicker than for 
main sequence stars (see e.g. Murray et al.~\cite{murray}).
In this case, the metal-rich material should be uniformly distributed
by convective mixing over a large portion of the star, resulting in a
negligible photospheric chemical alteration even for rather large amounts
of accreted material.
Late accretion, when the star is approaching or has already reached the 
main sequence, is likely required to produce observable differences.
This may happen due to infall of residual planetesimal/asteroids.
The age distribution of the main sequence stars hosting debris disks
(Habing et al.~\cite{habing01}), 
and of the impact craters on Solar System solid 
bodies, indicate that a significant infall of gas-poor material may be
present until 300-400 Myr after star formation; at this epoch 
the convective zone of a solar-mass star has been thinned enough
that significant alterations of the surface abundances may be produced.
The ingestion of planets scattered toward the star by dynamical
interactions (Marzari \& Weidenschilling \cite{marzari02}) might also
produce metallicity enhancements at late phases.

The signatures of accretion should be more evident for
stars with smaller mixing zones,  and disappear for evolved stars, when
the outer convective region penetrates deeply into the star. 
The observations of lithium depletion for stars in the range 
$1.2-1.5~M_{\odot}$ (the so-called Li-dip, Boesgaard \& Tripicco 
\cite{boesgaard86}) indicate that an additional  mixing
mechanism is at work, other than standard convective 
mixing\footnote{Alternative explanations of Li-dip depletion such as mass
loss or gravitational sedimentation do not seem to be supported 
by observations, see Do Nascimento et al.~(\cite{donascimento}).}. 
It must be further noted that the lithium dip is not present in open clusters
as young as Pleiades (Pilachowski et al.~\cite{pilachowski}), 
indicating an age dependence. 
Such an extra-mixing mechanism, besides destroying lithium, 
should also efficiently flatten the radial abundance profiles of the star, 
if enrichment of external layers occurred. Therefore the
stars that should show the most prominent signatures of accretion are early 
G-dwarfs and late F-dwarfs, with masses smaller than those in the 
Li-dip (Murray et al.~\cite{murray}). 
The occurrence of 'Li-dip' extra-mixing might also help in explaining
the lack of F stars with very high metallicity ([Fe/H]$ > +0.50$), 
predicted by the pollution
scenario when considering only the extension of the convective zone.

The metal abundance of planet hosts does not appear to be correlated
with the thickness of the convective zone 
(Pinsonneault et al.~\cite{pinsonneault}, Santos et al.~\cite{santos04}).
This seems to indicate that pollution is not the most relevant
contribution to the high metallicity of stars with planets.
On the other hand, two possible effects might in principle 
explain such a lack of a trend as a function
of the thickness of the convective enevelope.
The first is the result of Vauclair (\cite{vauclair}), who suggests
that, when the metallicity difference between the inner and outer parts
of a star exceedes a threshold, further mixing is induced in the transition
zone, therefore diluting the atmospheric metallicty enhancement.
The second effect can be considered if the metal enrichment is
dominated by the ingestion of already formed planets. According to the 
hydrodynamical simulations by Sandquist et al.~(\cite{sandquist98}), 
a $1.22~M_{\odot}$ star should be able to
only partially consume an engulfed giant planet within its thin convective
zone. Depending on the actual distribution of heavy elements within the planet
(possibly mostly confined in the planet core), the resulting alteration of
chemical composition might be even smaller than for a solar-like star that 
is able
to consume nearly completely a Jupiter-like planet within its convective zone.

On the other hand, several observational results support the
occurrence of pollution phenomena.

Smith et al.~(\cite{smith}) found indications of systematic
element-to-element differences as a function of the condensation
temperature on some planet host stars. This could be explained by the infall
of material devoid of light elements, such as rocky planets
or asteroids. However, such chemical composition trends can be confused
with galactic chemical evolution trends, and can be more easily studied in
binaries and clusters.
Gratton et al.~(\cite{paper1}), hereafter Paper I, performed a 
differential analysis of 6 visual main sequence binaries.
They found one pair (HD~219542) with a 0.09 dex iron content difference,
four pairs with very similar composition and one ambiguous case (HD~200466). 
Laws \& Gonzalez (\cite{laws01}) also found a small difference
between the components of 16 Cyg.
A similar approach can be pursued to look for chemical anomalies
among open cluster stars. Paulson et al.~(\cite{paulson}) studied
55 FGK dwarfs in the Hyades. 
They found two stars with abundances 0.2 dex larger
than the cluster mean, but their membership is questionable.
The Pleiades were studied by Wilden et al.~(\cite{wilden});
they found one possible candidate out of 15 stars with abundance 0.1
dex above the cluster mean and very little scatter for the other stars
(0.02 dex).
No correlation appears to be present between iron and lithium abundances.

The presence of $^6$Li might be a strong indication of accretion phenomena,
since this element is easily destroyed and cannot be observed
in main sequence stars unless recently fallen onto the star
from an external source. The presence of $^6$Li in HD~82943 is controversial
(Reddy et al.~\cite{reddy}, Israelian et al.~\cite{isr03}). If confirmed,
this would indicate the infall of a giant planet.
Anomalies of the most abundant  $^7$Li isotope are instead easier to find
(see e.g. Pasquini et al.~\cite{pasquini_m67}, King et al.~\cite{king}, 
Martin et al.~\cite{martin02}) but they are more difficult to 
interpret as they might not be linked to planetary pollution.

Murray et al.~(\cite{murray}) found that the Sun
should have ingested some $2~M_{\oplus}$ of meteoritic material
(about $0.4~M_{\oplus}$ of iron), considering the drop of iron density 
in the asteroid region and the time distribution
of the impact craters. This corresponds to a metallicity 
enhancement of 0.017 dex.
Helioseismology currently does not allow a confirmation of the presence of
such a small metal abundance difference between the inner
and the outer regions (Winnick et al.~\cite{winnick02}), with upper limits
on the amount of accreted iron about four times higher than the Murray et 
al.~(\cite{murray}) estimate\footnote{The comparison between the results
quoted in the original papers show some inconsistencies because of 
different assumptions on the
composition of meteoritic material. In this paper, we adopt the meteoritic
abundances by Lodders (\cite{lodders}). The mass fraction of iron
is thus 18.3\% of the meteoritic material.}.

From this discussion, it appears that the comparison of the chemical
composition of wide binaries is a very powerful approach to study
the occurrence of planetary pollution.
The number of pairs studied in Paper I is too small to allow
significant inferences about the frequency and the amount of chemical anomalies
in visual binaries.
In this paper, we significantly enlarge  the number of pairs
studied (23), with various improvements in the analysis procedure.
We will then combine the results of the present analysis with the
literature results for Hyades and Pleaides,  to draw 
tentative conclusions about the impact of planetary material pollution.


\section{Observations and data reduction}

A radial velocity survey aimed to find planets around stars in wide
binaries is  on-going at TNG (the Italian Telescopio Nazionale Galileo on
La Palms, Canary Islands, Spain)  using the
high resolution spectrograph SARG (Gratton et al.~\cite{papersarg}).
The goals of the project are the study of the dynamical effects
due to the presence of the companion on the existence and orbital
characteristics of planetary systems, and the study of chemical
composition differences possibly caused by the ingestion of planetary material.
A recent report describing the status of the survey is given by
Gratton et al.~(\cite{lapalma}).

The radial velocities are determined using the iodine cell technique.
A high resolution, high quality spectrum of each star, free of iodine 
lines, is used as template for the radial velocity analysis.
Such observational material is fully suited for a careful abundance
analysis, without further observational efforts.
All the spectra were acquired with the Yellow Grism, that covers
the spectral range 4600-7900~\AA~without gaps, and the 0.25 arcsec slit.
The resulting resolution is  R=150000 (2 pixel sampling).
The insertion of a suitable filter avoids any grism second order
contamination in the red extreme of the spectra.
During the observations, the slit was usually oriented perpendicularly to
the separation of the components
to minimize the contamination of the spectra by the companion.

We present in this paper the stellar templates acquired 
between June 2001 and November 2002. 
We also reanalyzed the pairs included in Paper I to take into account the 
modifications in our analysis procedure.
The spectra were reduced in the standard way using IRAF\footnote{IRAF is 
distributed by 
the National Optical Observatory, which is operated by the Association
of Universities for Research in Astronomy, Inc., under contract with the
National Science Fundation}.
The signal to noise ratio per pixel in most cases exceeds 150.


\section{Abundance analysis}

The analysis procedure is similar to that performed in Paper I, with
some differences outlined in detail below.
The unidimensional spectra were analyzed using routines within
the ISA code (Gratton \cite{rosa}).
The continuum level was determined by fitting a cubic spline through
automatically selected regions of the spectra, excluding discrepant points
identified by eye.
Then the equivalent widths (EWs) were measured using an automatic procedure,
that estimates a local continuum level and perform a Gaussian fit to
the selected lines (see Bragaglia et al.~\cite{n6819} for 
details)\footnote{The 
equivalent widths will be made available in
electronic form.}. The assumption of a Gaussian line profile fails
for stars with rotational velocities larger than about 5-10 km/s
(at the very high resolution of our spectra the contribution of 
instrument profile becomes small). The resulting EWs are 
overestimated for these stars. We then apply a suitable correction, 
determined from spectral synthesis of stars with different rotational 
velocities, to the EWs of stars with $v \sin i$ larger than
5 km/s (HD~8071 A/B, BD+231978A/B, HD 108574/5, HD~216122A/B).

The line list is the same used in Paper I, with the addition of some
vanadium lines from Whaling et al. (\cite{whaling}).

\begin{figure}
   \includegraphics[width=9cm]{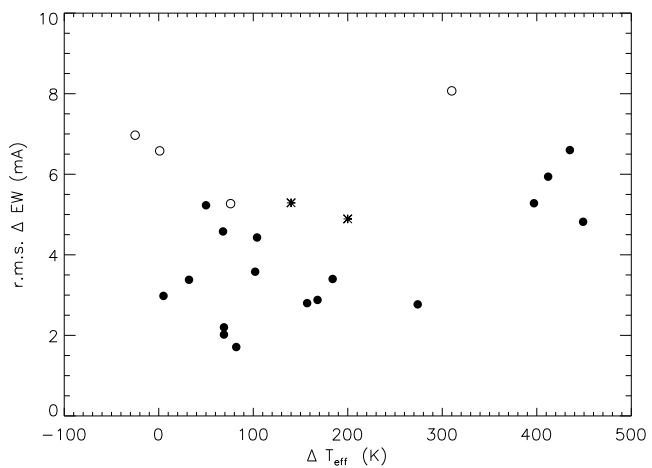}
      \caption{r.m.s scatter in equivalent width differences 
       $\Delta EW$ between the components
       of the binaries analyzed in this paper as a function of temperature
       difference $\Delta T_{eff}$. 
       Open circles represent stars with rotational broadening
       larger than about 5 km/s, filled circles slowly 
       rotating stars, asterisks the two pairs with slightly evolved 
       primaries.}
        
         \label{f:deltaew}
   \end{figure}

The errors of EWs can be estimated by the rms of the 
difference of EWs between the components of pairs whose
temperatures differ by less than 200~K, about
2-3 m\AA\ (Fig.~\ref{f:deltaew}).
Stars that have significant rotation have larger errors.
The increase of the rms of the difference for stars with larger
temperature difference and for the pairs whose primary appears
to be evolved out of the main sequence may be explained by the 
intrinsic variations of the line-by-line sensitivity to 
temperature and gravity changes.
The signal to noise ratio seems to play a  role in 
EW errors only 
for the few spectra with S/N lower than about 100-120.


The abundance analysis made use of Kurucz (\cite{kurucz95}) model atmospheres,
computed with the overshooting option switched off 
(Castelli et al.~\cite{castelli97}).
The primaries were analyzed differentialy with respect to the Sun,
then the line-by-line abundance differences between primaries and
secondaries were determined.

Improvements with respect to Paper I concern the inclusion of the damping
wings in the analysis of lines with $\log (EW/\lambda) > -5.4$ (in 
the previous analysis this limit was fixed to --5.1). 
The calculation of the damping constant is based on the formalism by
Barklem et al.~(\cite{barklem}), while in Paper I we used an enhancement factor
to the Unsold formula (Unsold \cite{unsold}).
These improvements allow a more appropriate
treatement of microturbulence (fixed to 0 in Paper I) and  
an extension toward stronger lines (up to 100 m\AA), while in Paper I
we limited ourselves to EWs $<50$~m\AA.
The line list is then expanded by about 50\%.
To remove outliers, we also used an automathic procedure 
that iteratively cleans the set of lines, disregarding outliers
that yield abundances differing more than 2.5 $\sigma$ from the
average abundance of the remaining lines,
while in Paper I outliers were removed by hand.

\subsection{Analysis of Primaries}

Table~\ref{t:mag} reports visual and absolute magnitudes, {\em B-V} colours,
parallaxes, masses and projected separations for the program stars. 
{\em V} band magnitudes are the
weighted average of Tycho and Hipparcos photometries (ESA \cite{hipparcos}), 
corrected to the standard
system using the calibrations by Bessell (\cite{bessell00}). {\em B-V} colours
are from Tycho and corrected using Bessell (\cite{bessell00}). The 
parallaxes are
from Hipparcos. When separate parallaxes were available for the two components
(HD~108574/5, HD~135101 A and B), a weighted average was performed.
The absolute magnitudes were corrected by the Lutz-Kelker effect following 
Hanson (\cite{hanson}). The differences with respect to Paper I concern
the inclusion of the Hipparcos photometry in the calculation of magnitude
differences, the correction to the standard system
of the Tycho photometry, the inclusion of some further literature measurements,
and the application of the Lutz-Kelker correction to parallaxes.

The stellar masses were determined from the isochrones of Girardi et
al.~(\cite{girardi02}), averaging the masses obtained for 1 Gyr 
main sequence stars having absolute magnitudes and temperatures equal 
to those of the program stars
for the appropriate metallicity\footnote{The periods of the observed systems
are too long for any reliable determination of dynamical masses (see
Desidera et al.~\cite{hd219542} for the case of HD 219542).}. 
To take into account the dependences
on temperature and metallicity, the masses were determined iteratively during
the abundance analysis. For the stars that show significant evolution out of
the main sequence (namely HD~135101A and HD~19440A), stellar masses were taken
from the suitable isochrone that fits the position of the components on the
color-magnitude diagram.
In Paper I the mass were derived using Tycho colours instead of the derived
temperatures (coupled with absolute magnitudes) and using 
Gray (\cite{gray}) tables (therefore neglecting metallicty effects).

 \begin{figure}
   \includegraphics[width=9cm]{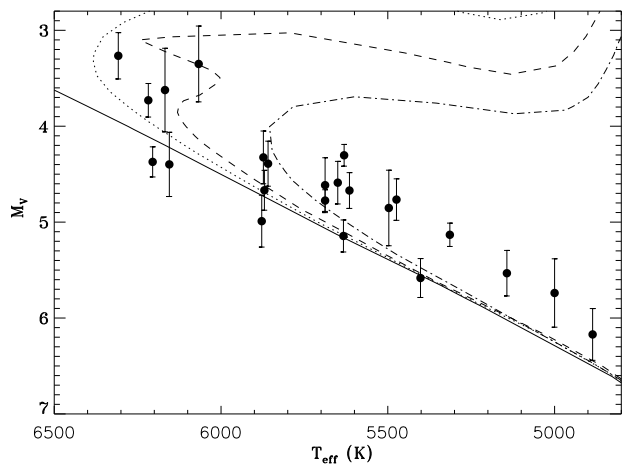}
      \caption{Color-Magnitude diagram for primaries using
         Hipparcos and Tycho magnitudes (corrected to the standard system using
         Bessell \cite{bessell00}) with Hipparcos parallaxes 
         and our spectroscopically determined effective temperatures. 
         The solar metallicity
         isochrones by Girardi et al.~(\cite{girardi02}) are overplotted 
         for ages of 1, 2.5, 4 and 8 Gyr.}
         \label{f:cmd}
   \end{figure}

\begin{table*}
   \caption[]{Visual and absolute magnitudes, B-V color, Hipparcos 
              parallaxes, stellar masses (derived iterativey during the
              abundance analysis using the spectroscopic
              temperatures and abundances) and projected binary separation 
              for program stars. As discussed in the text, the formal errors
              on magnitudes and colors are likely underestimated.}
     \label{t:mag}

\scriptsize{
      
       \begin{tabular}{lccccccccccc}
         \hline
         \noalign{\smallskip}
         Object      &  $V_{A}$ &  $V_{B}$   & $(B-V)_{A}$ & $(B-V)_{B}$  & $\pi$  & $M_{V(A)}$  & $M_{V(B)}$ & $M_{A}$ & $M_{B}$ & $\rho$ & $\rho$ \\
         &  &  & &  & mas  &  &  & $M_{\odot}$ & $M_{\odot}$ & $''$ & AU \\

         \noalign{\smallskip}
         \hline
         \noalign{\smallskip}

HD~8009    & $8.819\pm0.007$ & $9.724\pm0.017$ & $0.64\pm0.03$ & $0.82\pm0.07$ & 15.43$\pm$2.02 & 4.61$\pm$0.29 & 5.52$\pm$0.29 & 0.95 & 0.85 & 4.7 & 326 \\
HD~8071    & $7.312\pm0.003$ & $7.573\pm0.004$ & $0.58\pm0.01$ & $0.60\pm0.01$ & 19.67$\pm$1.58 & 3.73$\pm$0.17 & 3.99$\pm$0.17 & 1.25 & 1.22 & 2.2 & 115 \\
HD~9911    & $9.428\pm0.008$ & $9.448\pm0.008$ & $0.90\pm0.04$ & $0.89\pm0.04$ & 20.40$\pm$3.32 & 5.74$\pm$0.36 & 5.76$\pm$0.36 & 0.85 & 0.85 & 3.9 & 213 \\
HD 13357   & $8.180\pm0.006$ & $8.647\pm0.010$ & $0.67\pm0.03$ & $0.73\pm0.05$ & 20.41$\pm$1.75 & 4.67$\pm$0.19 & 5.14$\pm$0.19 & 0.98 & 0.93 & 5.9 & 297 \\
HD~17159   & $8.775\pm0.007$ & $8.923\pm0.007$ & $0.54\pm0.03$ & $0.53\pm0.03$ & 14.65$\pm$2.24 & 4.40$\pm$0.33 & 4.55$\pm$0.33 & 1.15 & 1.12 & 2.7 & 203 \\
HD~19440   & $7.874\pm0.004$ & $8.574\pm0.007$ & $0.47\pm0.01$ & $0.53\pm0.02$ & 12.55$\pm$1.39 & 3.27$\pm$0.24 & 3.96$\pm$0.24 & 1.25 & 1.14 & 6.0 & 501 \\
HD 30101   & $8.782\pm0.009$ & $8.848\pm0.009$ & $0.82\pm0.04$ & $0.91\pm0.03$ & 23.43$\pm$2.55 & 5.53$\pm$0.24 & 5.66$\pm$0.24 & 0.84 & 0.83 & 4.6 & 205 \\
HD 33334   & $8.023\pm0.004$ & $8.857\pm0.009$ & $0.70\pm0.03$ & $0.80\pm0.07$ & 21.39$\pm$2.18 & 4.59$\pm$0.22 & 5.42$\pm$0.22 & 1.00 & 0.89 & 4.7 & 229 \\
HD~66491   & $9.253\pm0.009$ & $9.312\pm0.009$ & $0.75\pm0.05$ & $0.67\pm0.05$ & 15.11$\pm$2.71 & 4.85$\pm$0.39 & 4.91$\pm$0.39 & 0.96 & 0.96 & 5.1 & 387 \\
BD+23 1978 & $9.395\pm0.010$ & $9.530\pm0.011$ & $0.80\pm0.06$ & $0.75\pm0.05$ & 24.04$\pm$2.97 & 6.17$\pm$0.27 & 6.31$\pm$0.27 & 0.84 & 0.84 & 2.6 & 115 \\ 
HD 106515  & $7.960\pm0.005$ & $8.234\pm0.007$ & $0.79\pm0.02$ & $0.83\pm0.03$ & 27.50$\pm$1.54 & 5.13$\pm$0.12 & 5.41$\pm$0.12 & 0.93 & 0.89 & 7.0 & 257 \\
HD 108574/5& $7.418\pm0.005$ & $7.972\pm0.007$ & $0.56\pm0.01$ & $0.67\pm0.02$ & 25.06$\pm$1.81 & 4.37$\pm$0.16 & 4.93$\pm$0.16 & 1.22 & 1.11 & 9.8 & 399 \\
HD 123963  & $8.758\pm0.006$ & $9.511\pm0.010$ & $0.60\pm0.03$ & $0.54\pm0.04$ & 13.81$\pm$1.74 & 4.33$\pm$0.28 & 5.08$\pm$0.28 & 1.08 & 0.96 & 2.6 & 200 \\
HD 132563  & $8.966\pm0.007$ & $9.472\pm0.010$ & $0.54\pm0.03$ & $0.57\pm0.05$ & 10.15$\pm$2.01 & 3.62$\pm$0.44 & 4.13$\pm$0.44 & 1.13 & 1.06 & 4.2 & 491 \\
HD 132844  & $9.022\pm0.006$ & $9.114\pm0.007$ & $0.55\pm0.05$ & $0.63\pm0.05$ & 16.57$\pm$2.05 & 4.99$\pm$0.27 & 5.08$\pm$0.27 & 1.01 & 0.99 & 3.0 & 192 \\
HD 135101  & $6.656\pm0.006$ & $7.500\pm0.009$ & $0.69\pm0.01$ & $0.75\pm0.02$ & 34.18$\pm$1.78 & 4.30$\pm$0.11 & 5.15$\pm$0.11 & 0.97 & 0.95 & 23.5 & 694 \\
HD 190042  & $8.755\pm0.009$ & $8.778\pm0.006$ & $0.73\pm0.03$ & $0.80\pm0.04$ & 16.52$\pm$1.64 & 4.76$\pm$0.22 & 4.79$\pm$0.22 & 0.96 & 0.94 & 4.2 & 264  \\ 
HD 200466  & $8.399\pm0.005$ & $8.528\pm0.006$ & $0.71\pm0.02$ & $0.79\pm0.03$ & 22.83$\pm$1.75 & 5.14$\pm$0.17 & 5.28$\pm$0.17 & 0.99 & 0.97 & 4.7 & 210 \\ 
HIP 104687 & $8.144\pm0.006$ & $8.189\pm0.006$ & $0.64\pm0.06$ & $0.71\pm0.06$ & 20.87$\pm$1.99 & 4.67$\pm$0.21 & 4.72$\pm$0.21 & 1.09 & 1.08 & 4.0 & 198 \\
HD 213013  & $8.982\pm0.008$ & $9.612\pm0.013$ & $0.81\pm0.04$ & $0.93\pm0.08$ & 21.59$\pm$2.01 & 5.58$\pm$0.20 & 6.21$\pm$0.20 & 0.93 & 0.84 & 5.5 & 263 \\
HD 215812  & $7.275\pm0.005$ & $7.576\pm0.006$ & $0.64\pm0.02$ & $0.70\pm0.02$ & 31.94$\pm$1.66 & 4.78$\pm$0.11 & 5.08$\pm$0.11 & 0.95 & 0.92 & 2.2 &  70 \\
HD 216122  & $8.062\pm0.011$ & $8.186\pm0.013$ & $0.58\pm0.02$ & $0.58\pm0.03$ & 13.11$\pm$2.36 & 3.35$\pm$0.40 & 3.47$\pm$0.40 & 1.23 & 1.22 & 5.3 & 464 \\
HD 219542  & $8.174\pm0.006$ & $8.547\pm0.008$ & $0.64\pm0.03$ & $0.71\pm0.04$ & 18.30$\pm$1.97 & 4.39$\pm$0.23 & 4.78$\pm$0.23 & 1.08 & 1.05 & 5.3 & 303 \\

         \noalign{\smallskip}
         \hline
      \end{tabular}

}

\end{table*}

Table \ref{t:magdiff} summarizes the magnitude differences for program stars,
including Hipparcos, Tycho and further ground-based photometry when available. 
Only literature observations reported in the standard system are included.
When available, we used the internal
errors as estimate of errors. 
A 0.04 mag error was assigned to the Docobo et al.~(\cite{docobo00}) 
magnitudes from the scatter of magnitude difference of  pairs with multiple
observations; a 0.02 mag error was assigned to the Simbad photometry
of HD~108575/5 and HD~135101 and to the Carney et al.~(\cite{carney94}) 
photometry of HD~200466.
The adopted magnitude difference is the weighted average of individual
determinations.
The resulting error is lower than 0.02 mag in most cases.

There is an offset of $0.052\pm0.016$ mag (rms 0.075 mag) 
between Hipparcos and Tycho magnitude differences.
The discrepancy is larger for pairs with a magnitude difference 
larger than 0.65 mag.
There is instead no significant offset between Hipparcos and the literature
determinations: the mean offset is $0.005\pm0.007$ mag (rms 0.023 mag).
The lower quality of Tycho photometry is also supported
by the very large difference of temperatures derived from {\em B-V} 
Tycho colors with those derived from spectroscopy (see Table~\ref{t:teff}).

The dispersion of Hipparcos and literature differences suggests that
quoted errors are underestimated by about 30\%. 
This may be due to the fact that in most cases the two components 
are not resolved by Hipparcos.
The Hipparcos epoch photometry (composite magnitudes) 
shows a scatter larger than internal errors in many cases 
(flag 'Duplicity induced variability').
Uncertainties in the transformation to the standard system, based
on Tycho colors, may also play a role.
We include such 30\% increase of errors in our adopted error bars.

\begin{table*}
   \caption[]{Magnitude difference for program stars.}
     \label{t:magdiff}
      
       \begin{tabular}{lccccl}
         \hline
         \noalign{\smallskip}
         Object      & $\Delta V$ & $\Delta V$ &  $\Delta V$ & Ref.  & $\Delta V$   \\
                     & Hipparcos  & Tycho      &  Other      & Other &  Adopted     \\
         \noalign{\smallskip}
         \hline
         \noalign{\smallskip}

HD~8009     & 0.935$\pm$0.021 & 0.763$\pm$0.046 & & & 0.905$\pm$0.065  \\ 
HD~8071     & 0.281$\pm$0.006 & 0.233$\pm$0.010 & & & 0.268$\pm$0.021  \\
HD~9911     & 0.026$\pm$0.013 &-0.023$\pm$0.033 & & & 0.019$\pm$0.017  \\ 
HD~13357    & 0.465$\pm$0.013 & 0.477$\pm$0.034 & 0.47$\pm$0.01 & 1 & 0.469$\pm$0.010  \\ 
HD~17159    & 0.148$\pm$0.011 & 0.166$\pm$0.028 & & & 0.150$\pm$0.010  \\ 
HD~19440    & 0.710$\pm$0.008 & 0.631$\pm$0.021 & $0.68\pm0.01$ & 1 &  0.692$\pm$0.016 \\ 
HD~30101    & 0.134$\pm$0.013 &-0.153$\pm$0.032 & 0.146$\pm$0.005 & 2 & 0.134$\pm$0.025 \\ 
            &                 &                 & 0.13$\pm$0.01 &  1 & \\
HD~33334    & 0.839$\pm$0.010 & 0.700$\pm$0.048 & 0.82$\pm$0.01 &  1 &  0.827$\pm$0.015 \\ 
HD~66491    & 0.057$\pm$0.013 & 0.077$\pm$0.047 &   & &  0.058$\pm$0.016  \\ 
BD+23 1978  & 0.139$\pm$0.016 & 0.092$\pm$0.041 &   & &  0.132$\pm$0.019  \\ 
HD~106515   & 0.273$\pm$0.009 & 0.263$\pm$0.024 &   & &  0.272$\pm$0.011  \\ 
HD~108574/5 & 0.554$\pm$0.012 & 0.572$\pm$0.015 & 0.57 & 3 & 0.563$\pm$0.011  \\ 
HD~123963   & 0.773$\pm$0.013 & 0.664$\pm$0.034 &    &  & 0.759$\pm$0.036  \\ 
HD~132563   & 0.511$\pm$0.013 & 0.451$\pm$0.039 &    &  & 0.505$\pm$0.018  \\ 
HD~132844   & 0.095$\pm$0.009 & 0.027$\pm$0.044 &    &  & 0.092$\pm$0.013  \\ 
HD~135101   & 0.894$\pm$0.020 & 0.833$\pm$0.012 & 0.85 & 3 & 0.849$\pm$0.017  \\  
HD~190042   & 0.021$\pm$0.008 & 0.037$\pm$0.031 &    & &  0.022$\pm$0.010   \\ 
HD~200466   & 0.133$\pm$0.008 & 0.101$\pm$0.021 & $0.15\pm0.01$ &  1 & 0.139$\pm$0.009  \\ 
            &                 &            & 0.17 &  4 & \\
HIP~104687  & 0.048$\pm$0.008 &-0.052$\pm$0.053 & 0.00 & 5 & 0.045$\pm$0.010  \\ 
            &                 &            & 0.06 &  5 & \\
HD~213013   & 0.631$\pm$0.016 & 0.627$\pm$0.047 & 0.62 & 5 & 0.628$\pm$0.017  \\ 
            &                 &            & 0.62 &   5 & \\
HD~215812   & 0.307$\pm$0.008 & 0.243$\pm$0.019 &     & &  0.297$\pm$0.022  \\ 
HD~216122   & 0.123$\pm$0.011 & 0.123$\pm$0.024 & $0.11\pm0.01$ & 1 & 0.117$\pm$0.009  \\ 
HD~219542   & 0.369$\pm$0.010 & 0.412$\pm$0.034 & 0.391$\pm$0.007 &  2 & 0.387$\pm$0.006 \\ 

         \noalign{\smallskip}
         \hline
      \end{tabular}

References: 1: Nakos et al.~(\cite{nakos95}); 
            2: Cuypers \& Seggewiss (\cite{cuypers99}); 
            3: Simbad; 
            4: Carney et al.~(\cite{carney94}); 
            5: Docobo et al.~(\cite{docobo00})

\end{table*}



As in Paper I, effective temperatures were determined from the 
ionization equilibrium.
This method gives robust values for the differences between the temperatures
of the two components, since in this case uncertainties in the absolute
magnitude (the largest contribution to errors) cancel out, the two components
being virtually at the same distance from the Sun.
Note however that errors are much larger for the analysis with respect to the
Sun (analysis of primaries), since in this case the full error bar in the
parallax must be taken into account.

In this paper, we also use vanadium lines, characterized by a temperature
sensitivity larger than typical of an iron line. 
The lack of a suitable number of
VII lines forced us to compare VI to FeII. Intrinsic differences of [V/Fe]
might in principle introduce spurious effects on our temperatures. However,
vanadium is not known to show peculiar relative abundance with respect to iron
for both Population I and Population II stars 
(Gratton \& Sneden \cite{gratton91}). 
Possible small intrinsic abundance differences due to chemical
evolution are not of concern in the differential analysis of the binary 
components. Spurious effects due to accretion of chemically fractionated 
material, more crucial for the goals of this work, are also not expected, 
since iron and vanadium have very similar condensation temperatures
(1334 and 1429~K respectively, according to Lodders \cite{lodders}).
The hyperfine splitting of V lines is included in our analysis
as in Gratton et al.~(\cite{large_field}). For the few lines for which
the hyperfine splitting constants are not available, a correction 
was applied considering the abundance difference of the other lines
due to  hyperfine splitting as a function of EW.

For the line lists we are using, the sensitivity of FeI-FeII and VI-FeII 
differences on temperature is about 0.0009 and 0.0015 dex/K respectively.
Temperatures from iron and vanadium were averaged.

Stellar gravities were calculated using the standard relation

\begin{equation}
\log g = {\rm const} + \log M + 0.4~(M_{V}+BC) + 4 \log T_{\rm eff}
\end{equation}

\noindent
where the constant term includes the solar quantities. We used the absolute
magnitudes from Hipparcos parallaxes, the stellar masses determined from the
isochrones of Girardi et al.~(\cite{girardi02}) and the bolometric corrections
by Kurucz (\cite{kurucz95}). Since the derived masses and 
gravities depend on temperatures
and metallicities, an iterative procedure is required.

The errors in the absolute magnitudes of the primaries are typically 
0.2-0.3 mag. These propagate into
errors of about 0.10 dex in $\log g$, therefore significantly affecting the
derived temperature from the ionization equilibrium ($\pm 30$~K) and the 
abundances for
the primaries ($\pm 0.04$~dex)\footnote{In the case of HD~132844A the
procedure used for the other stars gives very high gravities ($\log g=4.74$),
This is likely due to a large error in parallaxes. For this star only, the
analysis of the primary was purely spectroscopic, with the temperature derived
from the excitation equilibrium and gravity from the ionization equilibrium.}. 
However,
such sources of error are not of concern for the differential analysis between
primaries and secondaries, where they cancel out. Only errors in relative
photometry (less than 0.05 mag in all but one case)  play a role in this
case.

Microturbulent velocity was originally fixed by eliminating trends of Fe~I 
abundance with expected equivalent width (see Magain \cite{magain}). 
Residual trends with expected equivalent width are often present for cool
stars ($T_{\rm eff}<$5300~K), even when the microturbulent velocity was set at
zero. These residual trends may be due to various causes, the most likely
being an incorrect (too large) temperature gradient of the adopted model
atmospheres. To reduce the impact of this effect on our differential
abundance analyses, in these cases we only considered weak lines, that form
deeper in the atmosphere.

\begin{figure}
   \includegraphics[width=9cm]{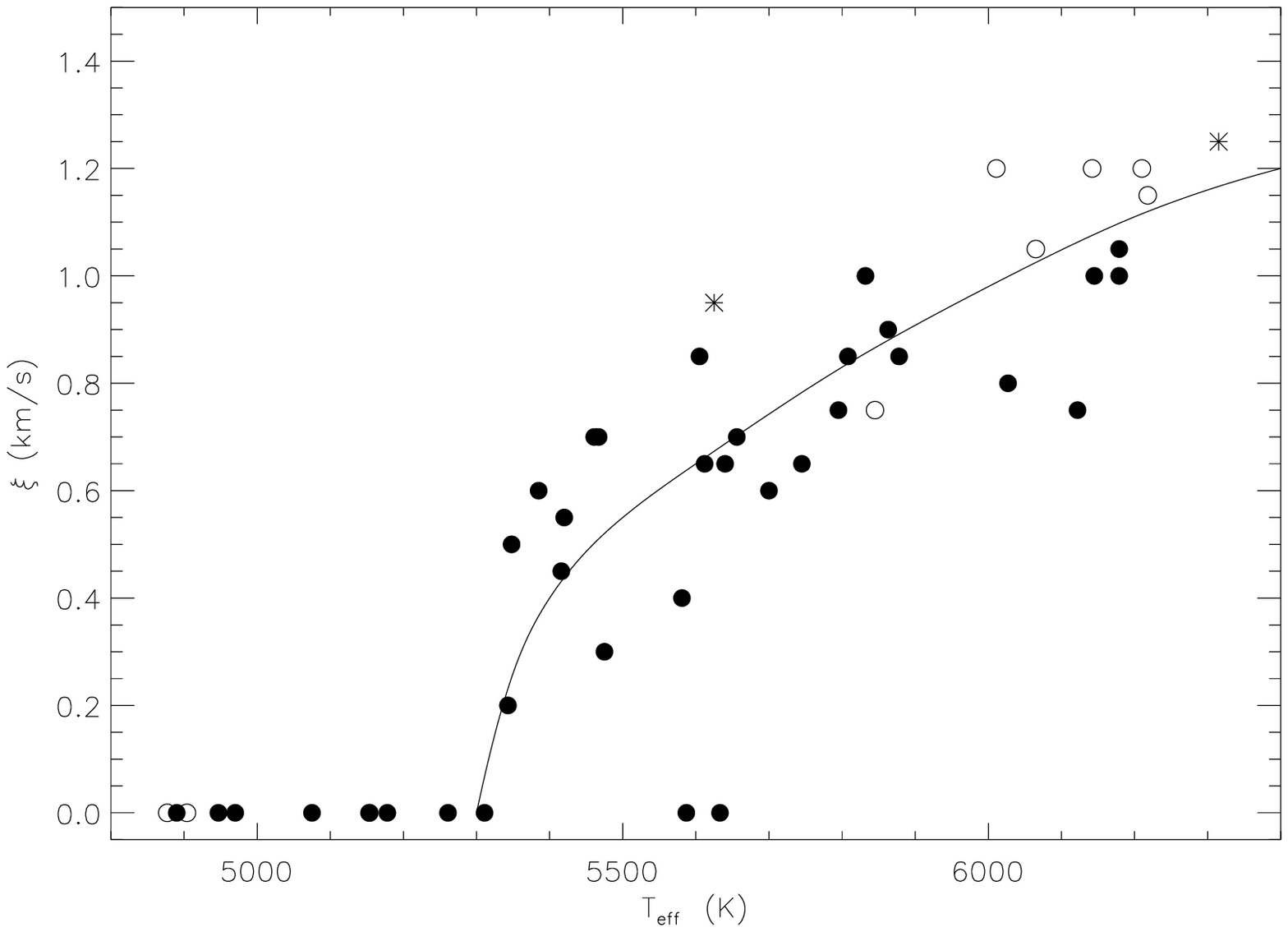}
      \caption{Microturbulence as a function of effective temperature
       resulting from the analysis performed leaving the microturbulence as a free
       parameter. Symbols as in Fig.~\ref{f:deltaew}.
       The relation between temperature and microturbulence adopted 
       in the analysis is shown as a continuous line.}
         \label{f:micro}
   \end{figure}

As expected for main sequence stars, there is a relation between
the temperature and the microturbulence derived as a free parameter
from the analysis (Fig.~\ref{f:micro}).
The residuals with respect to a global fitting to the
temperature-microturbulence relation (excluding rotating stars
and the two slightly evolved stars) show a dispersion
of 0.22 km/s. This is comparable to the errors in determination of 
individual values of microturbulence, estimated to be about 0.2~km/s.
Since the scatter around the fitting relation appears to be 
dominated by measurement errors, we adopted the values given
by the relation (shown in Fig.~\ref{f:micro}), iterating the process if 
the analysis with the revised microturbulence values requires
some temperature and abundance changes.

For the evolved stars HD~135101A and HD~19440A we instead adopted the
values of microturbulence resulting from the stellar analysis (as 
expected they are larger than the standard 
relation by 0.25 and 0.10 km/s respectively).

Table~\ref{t:atm_param} summarizes the atmospheric parameters adopted 
in the abundance analysis.
Table \ref{t:primaries} shows the results of the analysis of the
primaries.

\begin{table*}
   \caption[]{Adopted atmospheric parameters.}
     \label{t:atm_param}
      
       \begin{tabular}{cccccccccccc}
         \hline
         \noalign{\smallskip}
         Object      &  $T_{eff(A)}$ &  $T_{eff(B)}$   & $\log g_{A}$ & $\log g_{B}$  & [A/H]$_{A}$ & [A/H]$_{B}$ & $\xi_{A}$  & $\xi_{B}$  \\
         \noalign{\smallskip}
         \hline
         \noalign{\smallskip}

HD~8009      & 5688 & 5291 & 4.30 & 4.45 & --0.20 & --0.14 & 0.75 & 0.00 \\ 
HD~8071      & 6218 & 6142 & 4.24 & 4.32 &  +0.26 &  +0.26 & 1.10 & 1.10 \\
HD~9911      & 5000 & 4968 & 4.40 & 4.39 &  +0.06 &  +0.01 & 0.00 & 0.00 \\
HD~13357     & 5615 & 5341 & 4.31 & 4.36 & --0.02 & --0.01 & 0.65 & 0.25 \\
HD~17159     & 6155 & 6051 & 4.46 & 4.48 &  +0.02 &  +0.02 & 1.10 & 1.00 \\
HD~19440     & 6308 & 6108 & 4.09 & 4.27 & --0.05 & --0.05 & 1.25 & 1.05 \\
HD~30101     & 5143 & 5061 & 4.39 & 4.39 & --0.14 & --0.14 & 0.00 & 0.00 \\
HD~33334     & 5650 & 5201 & 4.30 & 4.39 &  +0.02 &  +0.03 & 0.70 & 0.00 \\
HD~66491     & 5497 & 5492 & 4.33 & 4.35 &  +0.03 &  +0.02 & 0.55 & 0.55 \\
BD+23 1978   & 4886 & 4911 & 4.51 & 4.58 &  +0.22 &  +0.22 & 0.00 & 0.00 \\
HD~106515    & 5314 & 5157 & 4.35 & 4.37 &  +0.08 &  +0.07 & 0.10 & 0.00 \\
HD~108574/5  & 6205 & 5895 & 4.49 & 4.55 &  +0.23 &  +0.23 & 1.10 & 0.90 \\
HD~123963    & 5873 & 5438 & 4.31 & 4.39 &  +0.11 &  +0.14 & 0.90 & 0.45 \\
HD~132563    & 6168 & 5985 & 4.15 & 4.27 & --0.18 & --0.19 & 1.10 & 0.95 \\
HD~132844    & 5878 & 5809 & 4.12 & 4.13 & --0.18 & --0.18 & 0.90 & 0.85 \\
HD~135101    & 5631 & 5491 & 4.17 & 4.44 &  +0.05 &  +0.05 & 0.95 & 0.55 \\
HD~190042    & 5474 & 5406 & 4.28 & 4.26 &  +0.02 &  +0.02 & 0.50 & 0.40 \\
HD~200466    & 5633 & 5583 & 4.51 & 4.54 &  +0.06 &  +0.04 & 0.70 & 0.65 \\
HIP~104687   & 5870 & 5801 & 4.45 & 4.44 &  +0.11 &  +0.12 & 0.90 & 0.85 \\
HD~213013    & 5402 & 4990 & 4.56 & 4.58 &  +0.08 &  +0.07 & 0.40 & 0.00 \\
HD~215812    & 5688 & 5586 & 4.37 & 4.43 & --0.19 & --0.21 & 0.75 & 0.65 \\
HD~216122    & 6067 & 6066 & 4.04 & 4.09 &  +0.25 &  +0.28 & 1.05 & 1.05 \\
HD~219542    & 5859 & 5691 & 4.34 & 4.42 &  +0.14 &  +0.14 & 0.90 & 0.75 \\

         \noalign{\smallskip}
         \hline
      \end{tabular}

\end{table*}

\begin{table*}
   \caption[]{Results of abundance analysis for primaries.}
     \label{t:primaries}
      
       \begin{tabular}{lccccccccccccc}
         \hline
         \noalign{\smallskip}
         Object      &  [Fe/H]I &  rms  & $N_{lines}$   & [Fe/H]II & rms & $N_{lines}$ & [V/H]I & rms & $N_{lines}$ \\
         \noalign{\smallskip}
         \hline
         \noalign{\smallskip}

HD~8009     & --0.195  & 0.084 & 126 & --0.194 & 0.164 &  32 & --0.195 & 0.113 & 28\\
HD~8071     &  +0.267  & 0.133 & 133 &  +0.203 & 0.197 &  30 &  +0.085 & 0.117 & 12\\
HD~9911     &  +0.058  & 0.082 &  73 &  +0.099 & 0.151 &  20 &  +0.177 & 0.162 & 27\\
HD~13357    & --0.012  & 0.130 & 146 & --0.045 & 0.094 &  28 & --0.109 & 0.090 & 32\\
HD~17159    &  +0.017  & 0.104 & 126 &  +0.038 & 0.129 &  33 &  +0.068 & 0.165 & 20\\
HD~19440    & --0.058  & 0.111 & 138 & --0.038 & 0.152 &  34 & --0.003 & 0.158 & 22\\
HD~30101    & --0.148  & 0.079 & 104 & --0.174 & 0.095 &  26 & --0.229 & 0.083 & 25\\
HD~33334    &  +0.015  & 0.078 & 122 &  +0.012 & 0.174 &  34 &  +0.013 & 0.108 & 32\\
HD~66491    &  +0.031  & 0.079 & 125 &  +0.016 & 0.130 &  25 & --0.013 & 0.118 & 30\\
BD+23 1978  &  +0.221  & 0.171 &  60 &  +0.282 & 0.157 &  14 &  +0.367 & 0.150 & 11\\
HD~106515   &  +0.078  & 0.065 &  85 &  +0.094 & 0.090 &  25 &  +0.120 & 0.092 & 23\\
HD~108574/5 &  +0.214  & 0.120 & 128 &  +0.259 & 0.198 &  28 &  +0.332 & 0.218 & 24\\
HD~123963   &  +0.111  & 0.086 & 140 &  +0.075 & 0.102 &  33 &  +0.010 & 0.091 & 28\\
HD~132563   & --0.185  & 0.103 & 139 & --0.186 & 0.109 &  34 & --0.197 & 0.173 & 19\\
HD~132844   & --0.182  & 0.077 & 130 & --0.208 & 0.153 &  32 & --0.249 & 0.059 & 23\\
HD~135101   &  +0.068  & 0.073 & 136 &  +0.077 & 0.125 &  31 &  +0.094 & 0.121 & 31\\ 
HD~190042   &  +0.045  & 0.092 & 130 &  +0.036 & 0.074 &  23 &  +0.019 & 0.143 & 31\\
HD~200466   &  +0.050  & 0.061 &  82 &  +0.055 & 0.137 &  30 &  +0.060 & 0.084 & 29\\
HIP~104687  &  +0.123  & 0.083 & 126 &  +0.106 & 0.157 &  31 &  +0.075 & 0.103 & 26\\
HD~213013   &  +0.084  & 0.088 &  84 &  +0.112 & 0.144 &  26 &  +0.151 & 0.129 & 30\\
HD~215812   & --0.195  & 0.083 & 136 & --0.198 & 0.074 &  29 & --0.204 & 0.094 & 27\\
HD~216122   &  +0.263  & 0.116 & 124 &  +0.263 & 0.141 &  23 &  +0.269 & 0.159 & 20\\
HD~219542   &  +0.135  & 0.084 & 132 &  +0.122 & 0.109 &  27 &  +0.102 & 0.143 & 28\\

         \noalign{\smallskip}
         \hline
      \end{tabular}

\end{table*}

\subsection{Differential analysis of secondaries}

The secondary of each pair was analyzed differentially with respect to the
primary, using strictly the same line set. The initial line list for the
differential analysis includes the lines not removed as outliers in the 
analysis of both components. Further iterative clipping is performed to exclude
lines that gives abundance differences outside $\pm 2.5 \sigma$ with
respect to the average of the other lines.

\begin{figure}
   \includegraphics[width=9cm]{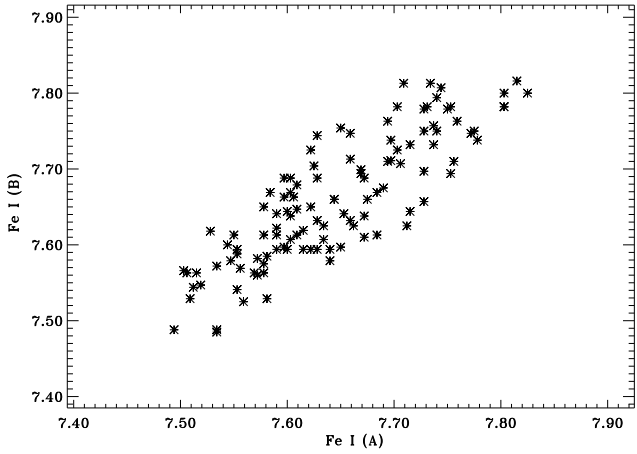}
      \caption{Iron abundance derived for each line for the components
               of HIP~104687~A and B. A clear correlation is present,
               indicating that the use of a line-by-line differential analysis
               significantly reduces the errors on abundance difference
               between the components.}
         \label{f:fe_lines}
   \end{figure}

The differential approach allows us to remove a number of
possible systematic errors in the analysis, as indicated by the internal
scatter of abundance difference that is lower than expected if the errors of
the two components were fully independent (Fig.~\ref{f:fe_lines}). The degree
of correlation between the abundances obtained from individual lines on the
spectra of the two components is a function of the difference of the
temperature of the stars: we found that the correlation is negligible when
this difference is larger than about 300~K. This is likely due to the 
combination of two factors: (i) for large temperature differences, line 
strengths are very different, due to the different impact of continuum
opacity; and (ii) systematic errors in the determination of the correct
continuum level become quite different. It is clear that the 
differential
analysis yields most accurate results when the two components are very similar.

Table~\ref{t:teff} shows the temperature differences given by different
methods. 

\begin{figure}
   \includegraphics[width=9cm]{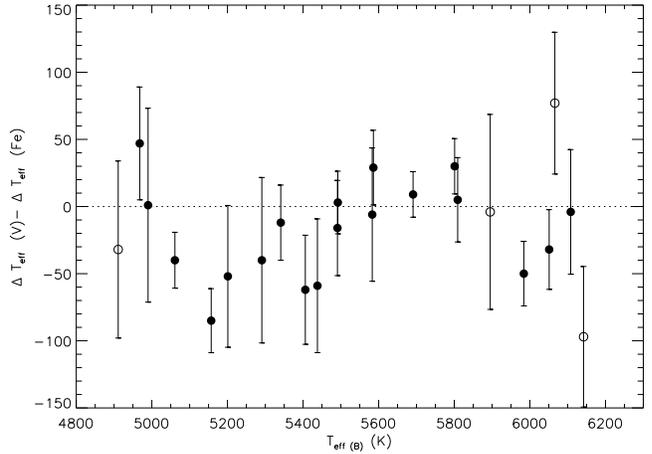}
      \caption{Differences between the temperatures derived using the iron and
               vanadium lines as a function of the temperature of the secondary.}
         \label{f:teff_feva}
   \end{figure}

The first and second column give the temperature difference from the
ionization equilibrium of iron and vanadium respectively and the third
column their weighted average.
As shown in Fig.~\ref{f:teff_feva}, for the pairs with temperature of
the secondary larger than 5450~K and excluding the rotating stars
characterized by much larger error bars, iron and vanadium temperatures
agree very well (mean offset $-3\pm7$~K, rms 25~K, 10 pairs).
A possible trend in the sense of vanadium giving larger temperatures
differences
may be present for the stars in the range 5100-5450~K, while
the situation is less clear for even cooler stars. 
Systematic trends in cooler stars may be caused by some residual
effects of hyperfine splitting or to the fact that most of the
lines in our line list become quite strong in the spectra of these
stars, for which a trend of abundance as a function of 
EWs still remains for such cool stars, as discussed above.
A better evaluation of the existence of systematic trends is
postponed to the analysis of further stars for which spectra
already have been obtained.

The excitation temperatures show an offset of $24\pm14$~K (rms 59~K)
for the 19 slow rotating pairs with respect to temperatures based
on ionization equlibrium.
The residual slopes of abundance difference as a function of 
excitation potential are correlated with those as a function
of equivalent width, suggesting a common origin (likely the
inadequacies of the atmospheric models and/or a possible 
underestimate of equivalent widths of strong lines when using
a Gaussian fit).
Therefore, the internal errors on the excitation temperature 
should not represent the full error bar.

Temperature differences from colors are calculated using the
calibrations by Alonso et al.~(\cite{alonso}). B-V colors are taken from
Table~\ref{t:mag}. V-I of HD~30101 and HD~219542 are from Cuypers \& Seggewiss
(\cite{cuypers99}) and those of HD~108584/5 and HD~135101 are from the 
Hipparcos Catalog. Transformation to the Johnson photometric system was 
performed following Bessell (\cite{bessell83}). 2MASS photometry was 
transformed to the
TCS photometric system using the calibrations provided by 
Cutri et al.~(\cite{cutri03})
and those by  Alonso et al.~(\cite{alonso94}).
The only pair for which Str\"omgren photometry is
available for both components is HD~135101 (Hauck \& Mermilliod 
\cite{hauckmerm}). A temperature difference of 160~K is derived.
It appears that temperatures based on colors have typically very poor
precision, likely because of the photometric errors induced by the small
separation of the components. 

The temperature difference from $\Delta V$ is
estimated considering the slope of the main sequence on the 1 Gyr solar
metallicity isochrone. It does not include evolutionary effects and therefore
represents an upper limit to the actual temperature difference.

 \begin{figure}
   \includegraphics[width=9cm]{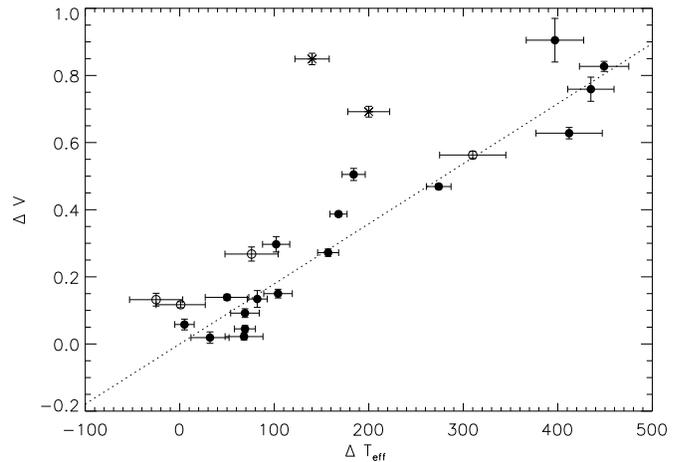}
      \caption{Magnitude vs temperature differences. The dotted line represents
               the typical slope alone the main sequence of a 1 Gyr solar 
               metallicity isochrone. Symbols as in Fig.~\ref{f:deltaew}.
               The two discrepant cases are
               HD~135101 and HD~19440, that have primaries clearly evolved off
               of the main sequence.}
         \label{f:dteff_dv}
   \end{figure}

\begin{table*}
   \caption[]{Temperature difference using different methods.
The weighted average between the temperature difference based on the 
ionization equilibrium
of iron and vanadium  was adopted because of its higher accuracy.}
     \label{t:teff}
      
       \begin{tabular}{lrrrrrrrrrr}
         \hline
         \noalign{\smallskip}
         Object      &  Ion. Eq. & Ion. Eq. & Ion. Eq. & Exc. Eq. & B-V & V-I & V-K & J-K & J-H & $\Delta V$ \\
                     &     Fe    &    V     &   mean   &          &     &     &     &     &     &            \\
         \noalign{\smallskip}
         \hline
         \noalign{\smallskip}

HD~8009     & $371\pm50$ & $411\pm36$ & $397\pm29$ &  $408\pm93$ &  563 &     & 649 &     &     & $\le$503\\
HD~8071     & $ 24\pm38$ & $121\pm36$ &  $76\pm28$ &  $234\pm55$ &   75 &     &     &     &     & $\le$149\\
HD~9911     & $ 62\pm33$ &  $15\pm26$ &  $32\pm20$ &  $-11\pm60$ & --44 &     &  14 &--39 &--149&  $\le$11\\
HD~13357    & $267\pm23$ & $279\pm16$ & $274\pm13$ &  $279\pm38$ &  188 &     & 272 & 176 & 488 & $\le$260\\
HD~17159    & $ 87\pm21$ & $119\pm21$ & $104\pm15$ &   $17\pm49$ & --54 &     &     &     &     &  $\le$83\\
HD~19440    & $198\pm28$ & $202\pm37$ & $200\pm22$ &  $282\pm60$ &  239 &     & 339 & 290 & 214 & $\le$385\\
HD~30101    & $ 55\pm17$ &  $95\pm12$ &  $82\pm10$ &  $126\pm27$ &  235 & 123 &  34 &  13 &--26 &  $\le$73\\
HD~33334    & $418\pm42$ & $470\pm32$ & $449\pm26$ &  $520\pm82$ &  295 &     & 570 & 262 & 489 & $\le$460\\
HD~66491    & $  9\pm20$ &  $ 6\pm12$ &  $ 5\pm10$ &   $21\pm38$ &--257 &     &  76 & 277 & 347 &  $\le$32\\
BD+23 1978  & $-43\pm50$ & $-11\pm43$ & $-25\pm28$ &  $106\pm224$&--142 &     &     &     &     &  $\le$73\\
HD~106515   & $ 98\pm20$ & $183\pm13$ & $157\pm11$ &  $206\pm49$ &  106 &     & 174 & 208 &  90 & $\le$151\\
HD~108574/5 & $307\pm57$ & $311\pm45$ & $310\pm35$ &  $392\pm87$ &  412 & 322 & 281 & 343 & 215 & $\le$313\\
HD~123963   & $399\pm39$ & $458\pm31$ & $435\pm24$ &  $539\pm82$ & --73 &     &     &     &     & $\le$422\\
HD~132563   & $159\pm17$ & $209\pm17$ & $184\pm12$ &  $222\pm38$ &  108 &     &     &     &     & $\le$281\\
HD~132844   & $ 72\pm25$ &  $67\pm19$ &  $69\pm15$ &   $36\pm22$ &  308 &     &     &     &     &  $\le$51\\
HD~135101   & $131\pm27$ & $147\pm23$ & $140\pm18$ &  $211\pm60$ &  188 & 158 & 136 & 124 &--24 & $\le$472\\
HD~190042   & $ 35\pm32$ &  $97\pm25$ &  $68\pm20$ &   $35\pm60$ &  220 &     &  22 &--55 & 120 &  $\le$12\\
HD~200466   & $ 38\pm41$ &  $44\pm28$ &  $50\pm23$ &   $17\pm92$ &  244 &     & 161 & 236 & 245 &  $\le$77\\
HIP~104687  & $ 87\pm16$ &  $57\pm13$ &  $69\pm11$ &   $53\pm22$ &  257 &     &   6 &--163& 242 &  $\le$25\\
HD~213013   & $413\pm58$ & $412\pm35$ & $412\pm35$ &  $570\pm109$&  315 &     & 335 & 112 & 203 & $\le$349\\
HD~215812   & $121\pm22$ &  $92\pm19$ & $102\pm14$ &  $124\pm49$ &  200 &     &     &     &     & $\le$165\\
HD~216122   & $ 50\pm42$ & $-27\pm32$ &   $1\pm26$ &   $45\pm71$ & --10 &     & 237 & 437 & 257 &  $\le$65\\
HD~219542   & $173\pm12$ & $159\pm15$ & $168\pm9$  &  $206\pm27$ &  246 & 238 & 232 &  23 & 220 & $\le$212\\

         \noalign{\smallskip}
         \hline
      \end{tabular}
 
\end{table*}

When considering the
differential analysis between the two components, parallaxes no longer
play a role; the same applies to absolute errors on mass estimates. 
Only errors on
relative photometry are relevant. The errors on magnitude
difference are less than 0.05 mag in all but one case (HD~8009). 
A 0.05 mag error in magnitude difference produces errors for
relative gravities of about 0.02~dex and for 7~K on relative temperatures. 
The contribution of magnitude difference to the temperature error
is therefore very small
The contribution of line-to-line scatter is larger for all the pairs
(Table \ref{t:teff}).

Table \ref{t:sensitivity} shows the sensitivity of the abundance differences
to variations of the atmospheric parameters, changing them one-by-one.

\begin{table*}
   \caption[]{Sensitivity to abundance and temperature difference to changes
              of atmospheric parameters for the pair HD~219542}
     \label{t:sensitivity}

\scriptsize{
      
       \begin{tabular}{lccccccc}
         \hline
         \noalign{\smallskip}
         Changes    &  $\Delta$ FeI &  $\Delta$ FeII & $\Delta$ VI & $\Delta$ FeI-FeII & $\Delta$ VI-FeII &  $\Delta T_{eff}$ &
                                      $\Delta T_{eff}$ \\
         &  &  &  &  &  &  (Fe) & (V)  \\

         \noalign{\smallskip}
         \hline
         \noalign{\smallskip}

$\Delta T_{eff}$  +30 K    &--0.018$\pm$0.008 & +0.010$\pm$0.013 &--0.033$\pm$0.024 &--0.028$\pm$0.016 &--0.043$\pm$0.027 &--32$\pm$17&
                                                                                                              --29$\pm$18  \\
$\Delta T_{eff}$ --30 K    & +0.017$\pm$0.008 &--0.012$\pm$0.013 & +0.028$\pm$0.024 & +0.029$\pm$0.016 & +0.040$\pm$0.027 & +32$\pm$17&
                                                                                                               +27$\pm$18 \\
$\Delta \log g$ +0.05 dex  & +0.002$\pm$0.008 &--0.019$\pm$0.013 & +0.001$\pm$0.024 & +0.021$\pm$0.016 & +0.020$\pm$0.027 & +23$\pm$17&
                                                                                                               +14$\pm$18  \\
$\Delta \log g$ --0.05 dex &--0.008$\pm$0.008 & +0.020$\pm$0.013 & +0.000$\pm$0.024 &--0.028$\pm$0.016 &--0.020$\pm$0.027 &--31$\pm$17&
                                                                                                              --13$\pm$18  \\
$\Delta$ [A/H] +0.05 dex   &--0.006$\pm$0.008 &--0.016$\pm$0.013 & -0.001$\pm$0.024 & +0.010$\pm$0.016 & +0.015$\pm$0.027 & +11$\pm$17&
                                                                                                               +10$\pm$18  \\
$\Delta$ [A/H] --0.05 dex  & +0.005$\pm$0.008 & +0.016$\pm$0.013 & +0.001$\pm$0.024 &--0.011$\pm$0.016 &--0.015$\pm$0.027 &--13$\pm$17&
                                                                                                               --10$\pm$18  \\
$\Delta \xi $  +0.15 km/s  & +0.017$\pm$0.008 & +0.030$\pm$0.013 & +0.005$\pm$0.024 &--0.013$\pm$0.016 &--0.025$\pm$0.027 &--15$\pm$17&
                                                                                                            --17$\pm$18  \\
$\Delta \xi $ --0.15 km/s  &--0.019$\pm$0.008 &--0.030$\pm$0.013 & -0.008$\pm$0.024 & +0.009$\pm$0.016 & +0.000$\pm$0.027 & +11$\pm$13&
                                                                                                               +15$\pm$18 \\

         \noalign{\smallskip}
         \hline
      \end{tabular}
 
}

\end{table*}

Table \ref{t:iron_diff} shows the results of the differential analysis for Fe~I
and Fe~II and Table \ref{t:diffabu} summarizes the results of temperature
difference, metallicity and iron content differences (Fe~I from Table
\ref{t:iron_diff}). We also give the internal errors (due to the line-to-line
scatter) and the global errors that include also the contribution due to the
errors on temperature differences. Errors due to microturbulence are not 
included.
The last column of Table \ref{t:diffabu} lists the
estimated amount of iron accreted by the richer in metal component of each 
pair, derived in Sect.~\ref{s:results}.

\begin{table*}
   \caption[]{Results of differential analysis}
     \label{t:iron_diff}
      
       \begin{tabular}{lrrrrrrrrr}
         \hline
         \noalign{\smallskip}
         Object   & $\Delta$ [Fe/H]I & rms & $N_{lines}$ & $\Delta$ [Fe/H]II & rms & $N_{lines}$ & $\Delta$ [V/H]I & rms & $N_{lines}$\\
         \noalign{\smallskip}
         \hline
         \noalign{\smallskip}
HD~8009      & --0.067 & 0.131 &  85 & -0.090 & 0.212 & 25 & -0.111 & 0.152 & 22\\ 
HD~8071      &  +0.001 & 0.113 & 116 & -0.046 & 0.133 & 17 & -0.113 & 0.151 & 13\\ 
HD~9911      &  +0.051 & 0.072 &  71 & +0.078 & 0.104 & 13 & +0.104 & 0.108 & 18\\ 
HD~13357     & --0.016 & 0.065 & 117 & -0.022 & 0.105 & 27 & -0.028 & 0.072 & 26\\ 
HD~17159     &  +0.012 & 0.097 & 102 & -0.004 & 0.085 & 28 & -0.027 & 0.095 & 13\\ 
HD~19440     &  +0.001 & 0.117 & 106 & +0.002 & 0.113 & 25 & +0.005 & 0.202 & 16\\ 
HD~30101     & --0.003 & 0.036 &  83 & -0.028 & 0.067 & 20 & -0.048 & 0.040 & 17\\ 
HD~33334     & --0.015 & 0.105 &  78 & -0.042 & 0.194 & 29 & -0.074 & 0.161 & 26\\ 
HD~66491     &  +0.006 & 0.063 & 112 & +0.010 & 0.080 & 22 & +0.015 & 0.030 & 21\\ 
BD+23 1978   & --0.012 & 0.145 &  43 & -0.028 & 0.135 & 12 & -0.049 & 0.116 &  5\\ 
HD~106515    &  +0.018 & 0.060 &  75 & -0.036 & 0.075 & 20 & -0.075 & 0.041 & 17\\ 
HD~108574/5  &  +0.004 & 0.160 & 105 & +0.002 & 0.242 & 24 & +0.000 & 0.190 & 17\\ 
HD~123963    & --0.037 & 0.137 & 120 & -0.070 & 0.169 & 27 & -0.105 & 0.150 & 20\\ 
HD~132563    &  +0.012 & 0.090 & 117 & -0.011 & 0.070 & 30 & -0.049 & 0.079 & 13\\ 
HD~132844    &  -0.004 & 0.052 & 110 & -0.008 & 0.116 & 27 & -0.005 & 0.059 & 13\\ 
HD~135101    &  +0.004 & 0.116 & 121 & -0.004 & 0.125 & 30 & -0.013 & 0.119 & 24\\ 
HD~190042    &   0.000 & 0.098 & 118 & -0.030 & 0.142 & 26 & -0.058 & 0.129 & 26\\ 
HD~200466    &  +0.020 & 0.117 &  70 & +0.009 & 0.180 & 28 & +0.000 & 0.111 & 21\\ 
HIP~104687   &  -0.015 & 0.045 & 113 & +0.001 & 0.045 &  9 & +0.018 & 0.056 & 19\\ 
HD~213013    &  +0.013 & 0.123 &  71 & +0.013 & 0.175 & 12 & +0.012 & 0.154 & 15\\ 
HD~215812    &  +0.017 & 0.092 & 127 & +0.034 & 0.091 & 25 & +0.050 & 0.084 & 22\\ 
HD~216122    &  -0.035 & 0.145 & 108 & +0.009 & 0.166 & 22 & +0.051 & 0.123 & 14\\ 
HD~219542    &  +0.001 & 0.057 & 110 & +0.006 & 0.042 & 20 & +0.012 & 0.081 & 23\\ 

         \noalign{\smallskip}
         \hline
      \end{tabular}
\end{table*}

\begin{table*}
   \caption[]{Final abundance difference  (Fe I from Table \ref{t:iron_diff}) 
              and estimates of iron possibly accreted}
     \label{t:diffabu}
      
       \begin{tabular}{lrrrrrr}
         \hline
         \noalign{\smallskip}
         Object      &  $\Delta T_{\rm{eff(A)}}$ & [Fe/H](A)   & $\Delta$ [Fe/H] & int. err. & error  &  $\Delta M_{Fe}$ \\
                     &  K &    &  &  &  &  $M_{\oplus}$ \\

         \noalign{\smallskip}
         \hline
         \noalign{\smallskip}

HD~8009      &  397$\pm$30 & --0.20 & --0.067 & 0.014 & 0.029 & 2.83$\pm$1.19 \\ 
HD~8071      &   76$\pm$28 &  +0.27 &  +0.001 & 0.009 & 0.021 & 0.03$\pm$0.77 \\
HD~9911      &   32$\pm$20 &  +0.06 &  +0.051 & 0.010 & 0.019 & 4.63$\pm$1.69 \\
HD~13357     &  274$\pm$13 & --0.01 & --0.016 & 0.009 & 0.013 & 0.80$\pm$0.63 \\
HD~17159     &  104$\pm$15 &  +0.02 &  +0.012 & 0.010 & 0.016 & 0.20$\pm$0.28 \\
HD~19440     &  200$\pm$22 & --0.06 &  +0.001 & 0.010 & 0.022 & 0.01$\pm$0.30 \\
HD~30101     &   82$\pm$11 & --0.15 &  -0.003 & 0.005 & 0.010 & 0.16$\pm$0.54 \\
HD~33334     &  449$\pm$26 &  +0.02 &  -0.015 & 0.013 & 0.025 & 1.10$\pm$1.86 \\
HD~66491     &    6$\pm$10 &  +0.03 &  +0.006 & 0.009 & 0.011 & 0.27$\pm$0.48 \\
BD+23 1978   &  -25$\pm$28 &  +0.22 & --0.012 & 0.025 & 0.032 & 2.64$\pm$7.31 \\
HD~106515    &  157$\pm$11 &  +0.08 &  +0.018 & 0.006 & 0.012 & 1.21$\pm$0.78 \\
HD~108574/5  &  310$\pm$35 &  +0.21 &  +0.004 & 0.015 & 0.034 & 0.10$\pm$0.78 \\
HD~123963    &  435$\pm$25 &  +0.11 & --0.037 & 0.012 & 0.024 & 2.50$\pm$1.60 \\
HD~132563    &  184$\pm$12 & --0.19 &  +0.012 & 0.007 & 0.013 & 0.12$\pm$0.13 \\
HD~132844    &   69$\pm$15 & --0.18 & --0.004 & 0.004 & 0.014 & 0.06$\pm$0.20 \\
HD~135101    &  140$\pm$18 &  +0.07 &  +0.004 & 0.010 & 0.019 & 0.20$\pm$0.96 \\
HD~190042    &   68$\pm$20 &  +0.05 &   0.000 & 0.010 & 0.019 & 0.00$\pm$0.96 \\
HD~200466    &   50$\pm$23 &  +0.05 &  +0.020 & 0.012 & 0.024 & 0.80$\pm$0.97 \\
HIP~104687   &   69$\pm$11 &  +0.12 & --0.015 & 0.005 & 0.010 & 0.49$\pm$0.33 \\
HD~213013    &  412$\pm$35 &  +0.08 &  +0.013 & 0.012 & 0.033 & 0.86$\pm$2.26 \\
HD~215812    &  102$\pm$14 & --0.20 &  +0.017 & 0.008 & 0.015 & 0.31$\pm$0.27 \\
HD~216122    &    1$\pm$26 &  +0.26 &  -0.035 & 0.013 & 0.026 & 0.99$\pm$0.73 \\
HD~219542    &  168$\pm$9  &  +0.14 &  +0.001 & 0.006 & 0.009 & 0.03$\pm$0.31 \\

         \noalign{\smallskip}
         \hline
      \end{tabular}

\end{table*}

Our analysis procedure makes use of standard stellar models
in the derivation of stellar masses. However, if  the outer
part of a star is enriched in metals, the model should be
systematically wrong. Stellar models with an enriched convective zone
have been calculated by e.g. Ford et al.~(\cite{ford}) and 
Dotter \& Chaboyer (\cite{dotter}).
In the case of 51 Peg, Ford et al.~(\cite{ford}) found that for 
a metallicity difference of 0.21 dex, the mass decreases from 1.05
to 0.94~$M_{\odot}$, while for $\tau$ Boo Dotter \& Chaboyer (\cite{dotter})
found a mass decrease from 1.36 to 1.32~$M_{\odot}$ for the same 
abundance difference. 

We estimated the effects of studing a  polluted star with an abundance 
difference of 0.21 dex using standard models taking as a reference 
the result by Ford et al.~(\cite{ford}) that predicts a larger
effect and is computed for a star with a temperature more typical of
the values of our program stars.
For fixed magnitudes, the stellar gravity 
is overestimated by 0.048 dex. This in turn implies 
an overestimatation  of the temperature derived from the
ionization equilibrium by about 25 K (with some further effects of the derived
stellar gravity) and then a $\sim$ 0.015 dex overestimatation of the abundance.
The induced error is then less than  10\% of the real metallicity difference.
Since we have not found a pair with a large abundance difference, the
impact on our result is very small.


\subsection{Comparison with Paper I and other literature data}
\label{s:lit}

In Paper I we derived an abundance difference of 0.09 dex for the
components of HD~219542, with the primary being more metal rich.
The present analysis does not confirm this result.

As described above, we made several modifications of
our analysis procedure with respect to Paper I.
The continuum level and the EWs were re-measured on the same spectra of
Paper I to be fully consistent with the other stars studied in
this paper, but without significant differences in the EWs
with respect to Paper I.
We also compared our EWs with those of Sadakane et al.~(\cite{sadakane03}).
We obtain a mean difference of $-0.050\pm0.074$~m\AA~~and 
$-0.051\pm0.069$~m\AA~~with an rms of 3.65 and 3.36~m\AA~~(24 lines)
for HD~219542 A and B respectively.

We performed several tests to understand the origin of the
change in abundance and temperature differences with respect to Paper I,
introducing one by one the changes between Paper I and this paper,
and starting from both the final solution given here and in Paper I.
We were able to explain most of the differences in terms of the
modifications of the analysis procedure  introduced.

The most relevant contributions come from the adoption of
different microturbulences for the two stars and for 
the changes
in the formalism for the damping constant (0.02-0.03 dex each one).

The lines selection also plays a role: using the 
the set of lines of Paper I (limit about 50 m\AA) with the
technique used in this paper causes a temperature difference of 
about 30~K (in the sense of HD~219542B being colder) and then
an abundance difference of about 0.02 dex.

Some further minor differences (less than 0.01 dex) are due to the 
inclusion of vanadium (for this pair the temperature differences
based on iron and vanadium differ
by just 9~K) and to the changes in the mass determination
(inclusion of the metallicity dependence and exclusion of 
Tycho colors).

The input metallicity used 
in the analysis also plays a role: when an abundance difference
is present between the components, 
adopting the same abundance difference 
in the atmospheric parameters causes 
a further increase of the measured difference.
From Table \ref{t:sensitivity} we see that this 
effect amounts to about 10\% of the input difference 
(i.e. about 0.01 dex for the case of HD~219542 in Paper I).

We also compared our results with those of Sadakane et al.~(\cite{sadakane03}).
They used spectra with a higher signal to noise ratio 
than ours (about 500) but 
measured a lower number of spectral lines (46 FeI and 7 FeII lines 
respectively). They determined the atmospheric
parameters spectroscopically, fixing the effective temperature and gravity
from the excitation and  ionization equilibrium respectively. 
They found ($T_{eff}$,
$\log g$, $\xi$, [Fe/H]) = (5830,4.45,0.95,0.13) and (5600,4.40,0.90,0.08) for
HD~219542 A and B respectively. 
Our parameters for the primary are fairly similar to their values, while
their temperature differences is larger by 62~K.
However, we note that the statistical error quoted by Sadakane 
et al.~(\cite{sadakane03})
for the temperature difference is 42~K (30~K for each star).
Furthermore, we note some inconsistencies in the parameters they derived.
Their spectroscopic analysis gives gravities larger 
for the warmer and brighter components ($\Delta \log g (A-B)=0.05$~dex), 
contrary to expectation for normal main sequence stars 
($\Delta \log g (A-B)= -0.06$~dex from the 1 Gyr isochrone).

As a  check that the discrepancy is not due to EW measurements, 
we analyzed our data using the Sadakane et al.~(\cite{sadakane03})
atmospheric parameters, and found  very similar abundance differences
than obtained by them, 
and the Sadakane et al.~(\cite{sadakane03}) EW using our technique.
The latter test gives a temperature difference 27~K larger than ours
(195 vs 168~K) and an abundance difference of 0.012 dex.

While we have no indication that the HD~219542 parallax 
is inaccurate, we checked the effect on the differential
analysis of the adoption of different temperatures for the
primary. For differences of 100~K with respect to the adopted
values, we get effects on the abundance difference at the 0.01 dex
level, likely because the relations between $\log g$ and $\xi$
with $T_{eff}$ are not linear.

The temperature difference adopted in this paper is 
compatible with the position of the stars in the color-magnitude diagram.
The primary might be slightly evolved (about 0.1 mag with respect to the main
sequence).

Finally, we note that 
the analysis of HD~219542B published by Santos et al.~(\cite{santos04}) gives 
($T_{eff}$, $\log g$, $\xi$, [Fe/H]) = (5732,4.40,0.99,0.17),
compatible within errors with our estimates.

One may wonder why the results for the other pairs studied in Paper I
do not change as did the values for HD~219542
(note that for HD 200466 we use in this paper a new, higher quality spectrum).
We think that this is due to the fact that, with the exception of HD~8071,
charactrized by a small temperature difference and thus with little
effect on the differential analysis, HD~219542 A is the warmest of the
stars studied in Paper I. The effect of assumptions about microturbulence
(zero for both components in Paper I, different values here) is then 
largest for this star.

We conclude that the abundance difference between the components of HD~219542
derived in Paper I is likely due the combination of several factors,
some random and some systematic, that worked by chance all in the same 
direction. This signals that caution should be applied when considering
abundance differences between pairs of stars, even if they appear
to be much larger than the internal errors.

The only other pair for which high resolution spectroscopic analysis
is published in the literature is HD~135101, whose components
were recently studied individually 
(i.e. as single stars) by Heiter \& Luck (\cite{heiter03}). 
They found ($T_{eff}$, $\log g$, $\xi$,
[Fe/H]) = (5700,4.20,0.90,$0.06\pm0.04$) and (5550,4.24,0.40,$0.10\pm0.05$)
for HD~135101 A and B respectively.
These parameters are fairly similar to ours. From our analysis there is
a much larger gravity difference between the components, derived from
the magnitude difference (0.85 mag).


\section{Results}
\label{s:results}

\begin{figure}
   \includegraphics[width=9cm]{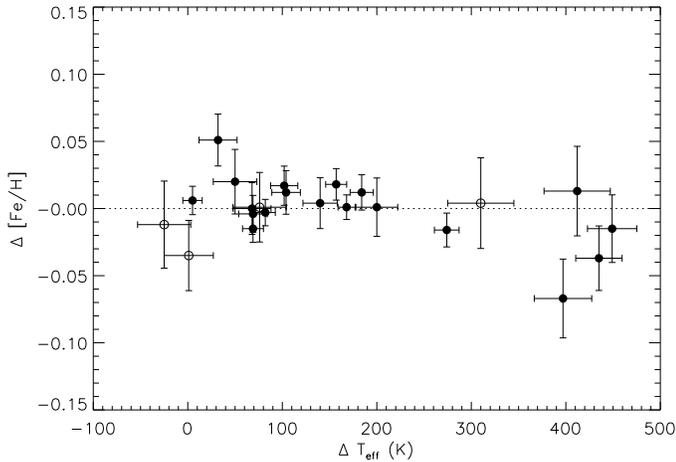}
      \caption{Iron abundance difference between the components of 
               pairs as a function of temperature difference. 
               Open circles represent stars with rotation broadening
               larger than about 5 km/s, filled circles slowly 
               rotating stars.}
         \label{f:metal_dteff}
   \end{figure}

\begin{figure}
   \includegraphics[width=9cm]{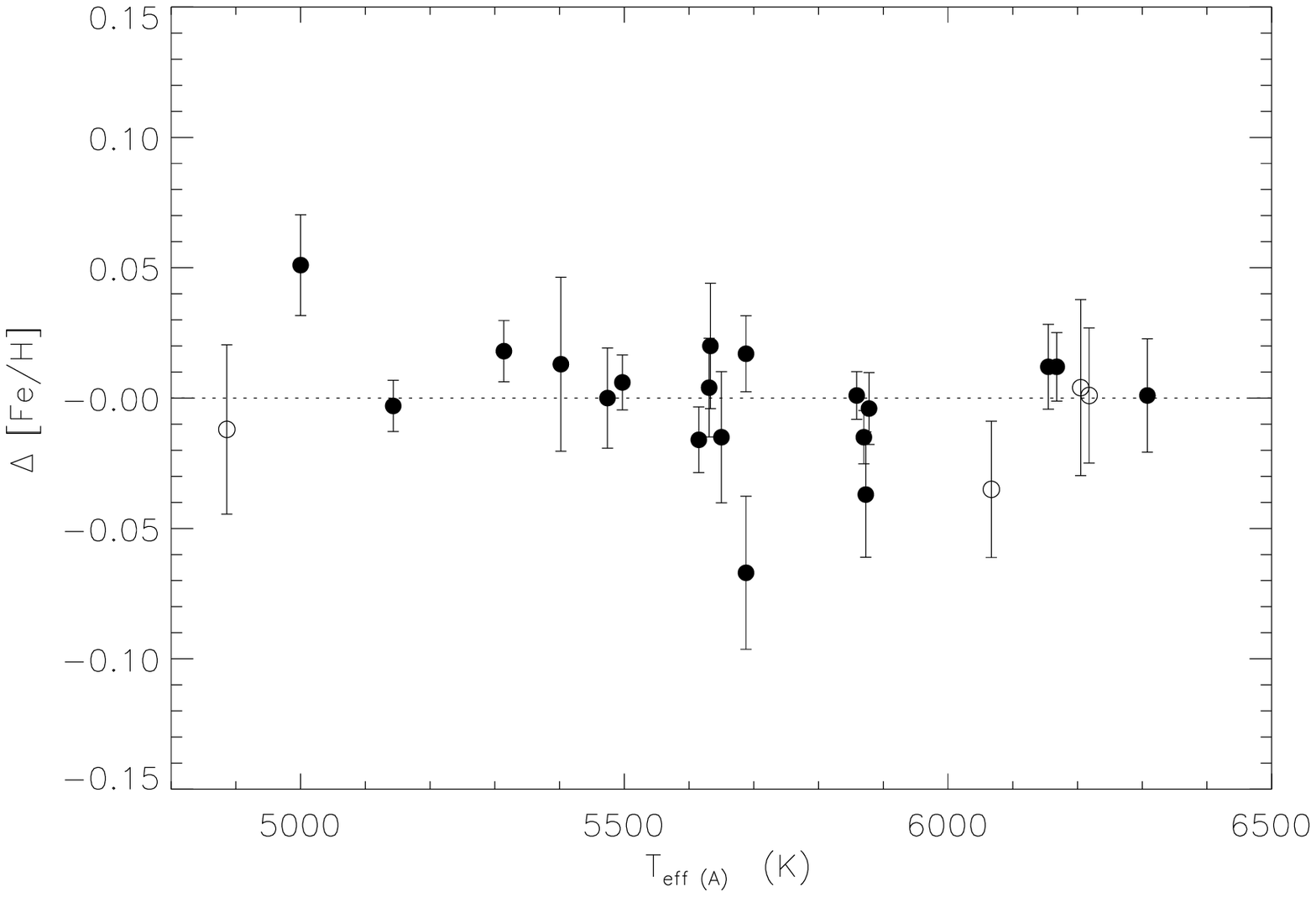}
      \caption{Iron abundance difference between the components of 
               pairs as a function effective temperature of the
               primary. Open and filled circles as in 
               Fig.~\ref{f:metal_dteff}.
}
         \label{f:metal_teff}
   \end{figure}

\begin{figure}
   \includegraphics[width=9cm]{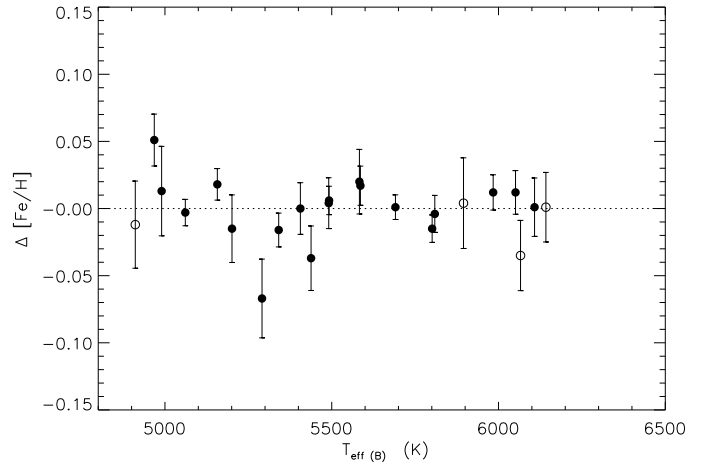}
      \caption{Iron abundance difference between the components of 
               pairs as a function effective temperature of the
               secondary. Open and filled circles as in 
               Fig.~\ref{f:metal_dteff}.
}
         \label{f:metal_teffb}
   \end{figure}

Fig.~\ref{f:metal_dteff}-\ref{f:metal_teffb} show the metallicity 
differences as a function of 
the temperature difference between the components, 
the temperature of the primary and that of the secondary.
The scatter of abundance difference is larger below 5500~K (Fig.~\ref{f:metal_teffb}). 
The pattern of $\Delta$ [Fe/H] as a function of the
temperature of the secondary rensembles
that of the difference of temperatures based on iron
and vanadium lines
but the correlation between $\Delta$ [Fe/H]  and
and $\Delta T_{eff} (Fe-V)$ is not statistically significant.

Four pairs have differences larger than 0.03 dex. 

The components of HD~8009 and HD~123963 have a fairly large temperature 
difference.
This causes errors larger than in most cases, making
the difference significant to about the 2.5 and 1.5 $ \sigma$ level 
respectively.
The magnitude difference between the components
is also quite uncertain for these pairs: the values given by Hipparcos
and Tycho differ by more than 0.1 mag (see Table~\ref{t:magdiff}).
The adoption of the Tycho magnitude difference for HD~8009 would imply a 
decrease of the abundance difference by about 0.01 dex.

The components of HD~9911 are quite cool (with only those of BD+23 1978
being even cooler). We saw that our analysis procedure has some
difficulty in handling such stars, likely because of inadequacies of
the atmosphere models, so  we think the the observed difference
(formally significant at about $2.5 \sigma$ level) is not real.

The components of HD~216122 show significant rotational broadening.
As for the other rotating stars, internal errors are larger,
making the observed difference of low significance. 
The star is also listed in the Hipparcos catalog as an unsolved variable.

Thus we have performed a differential abundance analysis of 23
wide binaries, with typical errors of about 0.02 dex.
The analysis appears more robust for the stars warmer than about
5450~K, while some inconsistencies appear to be present for
cooler stars. Such warmer stars are also the most
interesting for our science goal, because of their thinner
convective envelope.
While there are some pairs with a marginal abundance difference 
(0.03-0.07 dex), we take with caution such differences because
of possible problems in the analysis of these pairs
(errors due to the large temperature difference: HD~123963 and HD~8009;
systematic errors of abundance analysis at low temperature: HD~9911),
or large random errors, because of the fast rotational velocity
(HD~216122).

The analysis of other elements, the search for trends in abundance difference
with condensation temperature and the determination of the lithium
abundance  will be presented elsewhere.

\begin{figure}
   \includegraphics[width=9cm]{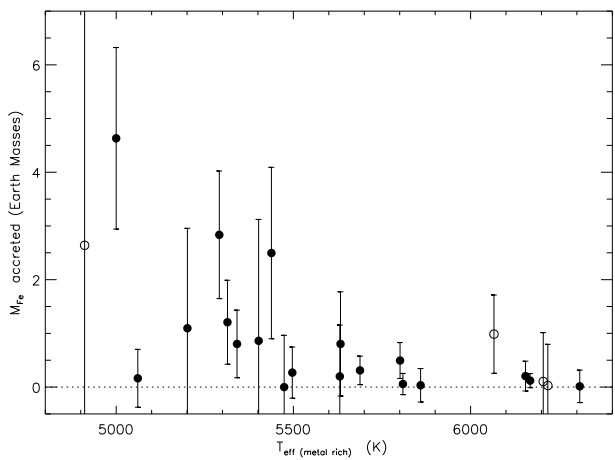}
      \caption{Estimate of iron accreted by the metal-rich component of each
               pair as a function of its effective temperature, taking into
               account the mass of the mixing zone as in Murray 
               et al.~(\cite{murray}). Open and filled circles as in 
               Fig.~\ref{f:metal_dteff}. The mass of meteoritic material
               is about 5.5 times the mass of iron.}
         \label{f:iron_accreted}
   \end{figure}

A large fraction of the pairs have abundance differences lower than 0.02 dex.
It is possible to estimate rough upper limits on metal-rich material accreted
by the stars after their convective zones have shrunk. We follow the
approach by Murray et al.~(\cite{murray}).
They consider the extension of the
convective zone during the main sequence as a function of stellar mass and
metallicity. Furthermore, they add a further mixing in the mass range
1.2-1.5~$M_{\odot}$, scaling the amount of mixing according to the observed
lithium depletion in this region\footnote{As noted in Sec.~\ref{s:intro},
the extra-mixing mechanism producing the Li dip has some dependence on 
stellar age. This is not taken into account in our estimate of the mass of 
the mixing zone.}. We can then calculate the amount of mass of
iron that, accreted on the (nominally) metal richer component of 
each pair, produces the
observed abundance difference (Fig.~\ref{f:iron_accreted}).

We are able to exclude at the 1~$\sigma$\ level the consumption
of 1~$M_{\oplus}$\
of iron for 10 out of 11  slow-rotating stars with stellar temperatures 
of both components higher than 5400~K. This amount of iron corresponds 
to less than 5~$M_{\oplus}$ of meteroritic material.
The accretion of a quantity of meteoritic material similar to that
expected to has been accreted from the Sun after the shrunking of its
convective zone (about 0.4~$M_{\oplus}$ of iron, Murray et 
al.~\cite{murray}) can be excluded at the 1 $\sigma$ confidence level
in five cases.

A more detailed estimate of the amount of iron accreted by stars of given
physical properties requires further investigation of the extension of the
mixing zone as well as suitable stellar models with enhanced metallicity in the
outer convective zone, that are beyond the scope of this paper.
The very small abundance difference of the warmest pairs may also be used
to constrain the diffusion of heavy elements within the star. 
From the models of Turcotte et al.~(\cite{turcotte}) we see that
the effect of iron diffusion at the end of the main sequence lifetime 
should be of 0.02 and 0.10 dex for a 
1.1 and 1.3~$M_{\odot}$ star respectively, i.e. within or close to
our detection limit.
Such investigations are postponed to a future paper, after the abundance
determination of further elements, very useful in constraining the 
diffusion scenario.

\section{Discusssion}

The frequency of planets is estimated to be about 8\% (Butler et
al.~\cite{butler_manchester}) and the stars hosting planets appear to have
metallicity larger than 0.25 dex with respect to normal field population stars 
(Laws et al.~\cite{laws03}). We have not found any pair among
23 with such large composition differences\footnote{Considering
the observed planet frequency, and  the differences of the 
mass of the convective zone between the components, it appears very unlikely
that both components of a binary system have such
metal rich material that their original composition
is alterated by a similar amount.}.
One star out of 15 in Pleiades
might have a metallicity enhancement of about 0.1 dex 
(Wilden et al.~\cite{wilden}), while two out of 55 in 
Hyades (with questionable membership) have metallicity difference of about 
0.20 dex with respect to the cluster mean (Paulson et al.~\cite{paulson}).

The fraction of stars with metal enrichment comparable to 
the observed metal overabundance of stars with 
planets appears to be lower than the fraction of stars with planets.
Taken at face value, this indicates that pollution phenomena {\it alone} 
cannot explain the
planet-metallicity connection. High metallicity then seems to play a 
role in favouring the presence of planets. Enrichment
phenomena might be present, as possibly indicated by  the still
controversial presence of $^6$Li in HD~82943, but they should be rare
and/or produce only a small composition alteration.
Increasing the statistics of wide binaries and stars in clusters
in the most suitable temperature range should allow one to constrain
the accretion scenario more tightly.

If we consider our sample and those of Pleiades and Hyades, we have
2 possible cases with a metallicity difference comparable to the typical
offset between planet hosts and normal field stars (the two Hyades
stars) out of 126 stars.
With the hypothesis that 8\% of these stars have planets (see below for
an important caveat) and that the $\sim0.25$ dex
metallicity offset is entirely due to planetary pollution, we obtain 
a probability of less than 1\% to observe just two cases out of 126 stars.
However, such a probability reaches 13\% if the actual planet frequency for
these stars is 4\%.

Some selection effect may be present considering that the
temperature and mass distribution of some of the samples studied
for the high-precision differential abundance analysis may be
different with respect to the field stars selected for radial velocity 
planet searches.

A more fundamental caveat to this discussion is that the components of 
wide  binaries considered in this study
(typical projected separations of about 100-400 AU, Table \ref{t:mag}) 
as well as stars in
open clusters like Pleiades and Hyades might have planet frequencies 
and pollution
histories different from the mostly single field star planets hosts. 
There are some hints that the properties of planets in binary systems are
different with respect to those orbiting single stars (Zucker \& Mazeh
\cite{zucker02}). This seems to indicate a significant role of orbital
interactions in the formation and evolution of planetary systems, whose
effects on the accretion of planetary material
should be considered. Dedicated searches for planets in various kind of
binaries (Desidera et al.~\cite{hd219542}; Eggenberger et
al.~\cite{eggenberger}) and in star clusters 
characterized by different dynamical conditions (see e.g. 
Gilliland et al.~\cite{gilliland}; Bruntt et al.~\cite{bruntt}; 
Street et al.~\cite{street}; Piotto et al.~\cite{p8})
are thus very useful to understand the effects of
the dynamical environment on the presence of planets and their orbital 
properties. This
will allow us also to determine whether the pollution properties derived by the
study of wide pairs and cluster stars can be extended to single stars 
(for which this is much more difficult to show).

\begin{acknowledgements}

   We thank the TNG staff for its help with the observations.
   This research has made use of the 
   SIMBAD database, operated at CDS, Strasbourg, France,
   and of data products from the Two Micron All Sky Survey,
   which is a joint project of the University of Massachusetts and the
   Infrared Processing and Analysis Center/California Institute of Technology,
   funded by the National Aeronautics and Space Administration and the
   National Science Foundation.

\end{acknowledgements}

\end{document}